\documentclass[english,aps,prb,amsmath,preprint,showpacs]{revtex4}
\usepackage[T1]{fontenc}
\usepackage[latin9]{inputenc}
\usepackage[a4paper]{geometry}
\usepackage{graphicx}
\usepackage{esint}

\makeatletter

\@ifundefined{textcolor}{}
{%
 \definecolor{BLACK}{gray}{0}
 \definecolor{WHITE}{gray}{1}
 \definecolor{RED}{rgb}{1,0,0}
 \definecolor{GREEN}{rgb}{0,1,0}
 \definecolor{BLUE}{rgb}{0,0,1}
 \definecolor{CYAN}{cmyk}{1,0,0,0}
 \definecolor{MAGENTA}{cmyk}{0,1,0,0}
 \definecolor{YELLOW}{cmyk}{0,0,1,0}
}

\input{tcilatex}
\usepackage{amssymb}
\usepackage{enumitem}
  \setenumerate{label={(\roman*)}}

\makeatother

\usepackage{babel}
\begin{document}

\title{Microwave-assisted switching of a nanomagnet: analytical determination
of the optimal microwave field}

\author{N. Barros, M. Rassam, and H. Kachkachi}

\affiliation{PROMES-CNRS and Université de Perpignan Via Domitia, 52 Avenue Paul
Alduy, 66860 Perpignan Cedex, France}

\begin{abstract}
We analytically determine the optimal microwave field that allows
for the magnetization reversal of a nanomagnet modeled as a macrospin.
This is done by minimizing the total injected energy. The results
are in good agreement with the fields obtained numerically using the
optimal control theory. For typical values of the damping parameter,
a weak microwave field is sufficient to induce switching through a
resonant process. The optimal field is orthogonal to the magnetization
direction at any time and modulated both in amplitude and frequency.
The dependence of the pulse shape on the applied field and damping
parameter is interpreted. The total injected energy is found to be
proportionnal to the energy barrier between the initial state and
the saddle point and to the damping parameter. This result may be
used as a means for probing the damping parameter in real nanoparticles. 
\end{abstract}
\maketitle

\section{Introduction}

Magnetic recording is a key technology in the field of high density
information storage. In order to increase thermal stability, small
nanoparticles with high anisotropy may be used. However, high fields
are then needed to reverse the magnetization but these are difficult
to achieve in current devices. To overcome this so called magnetic
recording trilemma, several solutions are being proposed. The mostly
investigated route, and which already leads to industrial applications,
is the heat-assisted magnetic recording\cite{thomson08}. It consists
in heating the particles by a laser which decreases the energy barrier
between the two energy minima and thereby the switching fields. However,
to avoid a loss of information, the heating must be very localized
and followed by a very fast cooling, and as such these devices must
be coupled to powerful heat dissipation systems. 

An alternative solution is to assist the switching by a microwave
(MW) field. In 2003 Thirion et al.\cite{thirion03} showed that the
combination of a DC applied field (static field) well below the switching
field with a small MW field pulse can reverse the magnetization of
a nanoparticle. Indeed, in the presence of a MW field with appropriate
amplitude and frequency, the magnetization precession synchronizes
with this field\cite{bertotti01,denisov06,bertotti09bis}. Then, energy
is pumped into the system thus allowing the magnetization to climb
up the energy barrier and cross the saddle point\cite{sun06,sun06bis,okamoto08,bertotti09,lyutyy09}.
Further experimental and theoretical studies proved that this process
is more efficient if the frequency of the MW field is sligthly lower
than the ferromagnetic resonance frequency of the nanoelements\cite{rivkin06,woltersdorf07,laval09}.
Moreover, the use of chirped MW fields was shown to be more efficient
to achieve switching\cite{rivkin06,sun06bis,okamoto08bis,wang09,cai12}.
This result is related to the anharmonicity of the energy well. Similar
results have been obtained in other areas of physics and chemistry,
like atomic or molecular spectroscopy\cite{meerson90,jewariya10}. 

In a previous work\cite{barros11} we developped a numerical method
based on optimal control theory which renders an exact solution for
the MW field that is necessary for the switching of a nanomagnet within
a given potential energy. The formulation of this method consists
in defining a cost functional and minimizing it using the conjugate
gradient technique. Our results confirmed that a weak MW field, modulated
both in amplitude and frequency, can induce the switching of the magnetization.
Furthermore, the injected energy has been found to increase with damping. 

The aim of the present study is to provide analytical expressions
and to compare them with our numerical results, by using simple energy
considerations. Moreover, the analytical developments presented here
confirm the effects observed numerically and provide clear interpretations
for the underlying physical processes. In this work, our investigations
are restricted to zero temperature, and as such the Landau-Lifshitz-Gilbert
equation is used to describe the magnetization trajectory. The additionnal
effects of thermal fluctuations will be the subject of a future study. 

In the first part, we analytically determine the optimal MW field
and demonstrate its dependence on the energy landscape (anisotropy,
applied field) and to the damping parameter. We then investigate the
trajectory of the magnetization in the presence of the optimal field
and show that it can be described by the Landau-Lifshitz-Gilbert equation
with a negative damping parameter. In the second part, the analytical
results are compared directly with the results obtained numerically
using the optimal control theory.

\section{Analytical calculation of the optimal microwave field\label{sec: Analytic-calculations}}

We consider a nanomagnet with spatially uniform magnetization which
can be modeled by a vector $\mathbf{M}=M_{S}\mathbf{m}$, where $M_{S}$
is the saturation magnetization and $\left\Vert \mathbf{m}\right\Vert =1$.
This nanomagnet is characterized by a given anisotropy (uniaxial,
biaxial, cubic...) and a damping parameter $\alpha$. In the presence
of a static magnetic field $\mathbf{H}_{0}$ lower than the Stoner-Wohlfarth
switching field, the potential energy surface presents several minima
separated by saddle points. 

At the initial time $t_{i}$, we assume that the magnetization is
in a minimum $\mathbf{M}_{i}=M_{S}\mathbf{m}_{i}$. Adding a microwave
(MW) field $\mathbf{H}(t)$ can then induce switching to another (target)
minimum $\mathbf{M}_{f}=M_{S}\mathbf{m}_{f}$. Our aim is to find
the optimal field $\mathbf{H}^{\mbox{opt}}(t)$ that achieves switching
in a given time $t_{f}$ while minimizing the energy injected into
the system. This criterion is relevant for experimental devices since
it amounts to reducing both the intensity and the duration of applied
fields and the subsequent heating of the system, which can be of interest
for magnetic recording or biomedical applications. This approach is
thus complementary to other theoretical studies which have focused
on the reduction of the switching time \cite{bertotti09,sukhov09}. 

For the sake of simplicity, we introduce the normalized fields $\mathbf{h}_{0}\equiv\mathbf{H}_{0}/H_{\mbox{an}}$
and $\mathbf{h}(t)\equiv\mathbf{H}(t)/H_{\mbox{an}}$, where $H_{\mbox{an}}\equiv2K/\mu_{0}M_{s}$
is the anisotropy field and $K$ the anisotropy constant of the nanomagnet.
We also define the normalized time $\tau\equiv\gamma H_{\mbox{an}}t$,
where $\gamma=1.76\times10^{11}\mbox{ (T.s)}^{-1}$ is the gyromagnetic
factor. For instance, for a cobalt particle of 3 nm diameter with
$K\approx2.2\times10^{5}\mbox{ J.m}^{-3}$ and $M_{s}\approx1.44\times10^{6}\mbox{ A.m}^{-1}$
we have $\mu_{0}\, H_{\mbox{an}}\approx305\mbox{ mT}$ and $t/\tau\approx1.86\times10^{-11}\mbox{ s}$.

\subsection{Energy and time trajectory in the presence of a microwave field\label{sub: energy and trajectory}}

If only the static field is applied, the energy density of the system
(divided by $2K$) reads $\mathcal{E}_{0}(\mathbf{m},\mathbf{h}_{0})=\mathcal{E}_{\mbox{an}}(\mathbf{m})-\mathbf{m}\cdot\mathbf{h}_{0}$
where $\mathcal{E}_{\mbox{an}}$ is the anisotropy energy density.
The normalized effective field is then defined by $\mathbf{h}_{\mbox{eff}}\equiv-\partial\mathcal{E}_{0}/\partial\mathbf{m}$.
If we add a MW field, the energy density becomes 
\begin{equation}
\mathcal{E}(\mathbf{m},\mathbf{h}_{0},\mathbf{h}(\tau))=\mathcal{E}_{\mbox{an}}(\mathbf{m})-\mathbf{m}\cdot\left(\mathbf{h}_{0}+\mathbf{h}(\tau)\right)\label{eq: w(m,h,b)}
\end{equation}
and the normalized total effective field now reads
\begin{equation}
\mathbf{\zeta}(\tau)\equiv-\frac{\partial\mathcal{E}}{\partial\mathbf{m}}=\mathbf{h}_{\mbox{eff}}+\mathbf{h}(\tau).\label{eq: zeta(t)}
\end{equation}

The time trajectory of the magnetization can be described by the driven
Landau-Lifshitz-Gilbert equation 
\begin{eqnarray}
\left(1+\alpha^{2}\right)\frac{d\mathbf{m}}{d\tau} & = & -\mathbf{m}\times\mathbf{\zeta}(\tau)-\alpha\mathbf{m}\times\left(\mathbf{m}\times\mathbf{\zeta}(\tau)\right).\label{eq: DLLE}
\end{eqnarray}

This allows us to express the energy variation of the system as follows
\begin{equation}
\frac{d\mathcal{E}}{d\tau}=-\mathbf{\zeta}(\tau)\cdot\frac{d\mathbf{m}}{d\tau}-\mathbf{m}\cdot\frac{d\mathbf{h}(\tau)}{d\tau}=-\mathbf{h}_{\mbox{eff}}\cdot\frac{d\mathbf{m}}{d\tau}-\frac{d}{d\tau}\left(\mathbf{m}\cdot\mathbf{h}(\tau)\right).\label{eq: dwdt}
\end{equation}

Next, we define the mobile frame $\left(\mathbf{m},\mathbf{u},\mathbf{v}\right)$
attached to the magnetization with $\mathbf{u}\equiv\mathbf{T}/T$
and $\mathbf{v}\equiv\mathbf{m}\times\mathbf{T}/T$, where $\mathbf{T}\equiv\mathbf{m}\times\mathbf{h}_{\mbox{eff}}$
and $T=\left\Vert \mathbf{m}\times\mathbf{h}_{\mbox{eff}}\right\Vert $.
The MW field can then be decomposed as $\mathbf{h}(\tau)=h_{m}(\tau)\mathbf{m}+h_{u}(\tau)\mathbf{u}+h_{v}(\tau)\mathbf{v}$.
In this frame, Eqs. (\ref{eq: DLLE}) and (\ref{eq: dwdt}) respectively
become
\begin{equation}
\left(1+\alpha^{2}\right)\frac{d\mathbf{m}}{d\tau}=\left(-1+\frac{\alpha\, h_{u}(\tau)+h_{v}(\tau)}{T}\right)\mathbf{T}-\left(\alpha+\frac{h_{u}(\tau)-\alpha h_{v}(\tau)}{T}\right)\left(\mathbf{m}\times\mathbf{T}\right),\label{eq: DLLE in mobile basis}
\end{equation}
\begin{eqnarray}
\frac{d\mathcal{E}}{d\tau} & = & \frac{-\alpha T-h_{u}(\tau)+\alpha h_{v}(\tau)}{1+\alpha^{2}}\, T-\frac{dh_{m}(\tau)}{d\tau}.\label{eq: dwdt in mobile basis}
\end{eqnarray}

We note that the parallel component of the MW field $h_{m}(\tau)$
has no direct effect on the magnetization trajectory and that only
its time derivative appears in the energy variation.

\subsection{Optimization of the MW field}

In order to find the optimal MW field fulfilling the requirements
described earlier we proceed in two steps. First, we define the critical
MW field which allows us to maintain the precession of the magnetization
by compensating the effects of damping (sec. \ref{sub: critical field}).
This field represents the lower limit for the optimal field sought.
Using this result, we find the optimal MW field minimizing the injected
energy (sec. \ref{sub: injected energy}) and check that it can induce
switching of the magnetization (sec. \ref{sub: trajectory}).

\subsubsection{MW field maintaining the precession: critical field \label{sub: critical field}}

In order to induce switching the MW field must at least compensate
for the effect of damping, which tends to take the magnetization back
to the initial equilibrium position. If the compensation is complete
the energy variation of the system $d\mathcal{E}/dt$ vanishes at
any time thus reflecting the conservation of energy. According to
Eq. (\ref{eq: dwdt in mobile basis}) an infinity of MW fields leads
to a full compensation of damping. For instance, any field that is
orthogonal to the magnetization so that $h_{m}(\tau)=0$ and satisfying
the equation $-h_{u}(\tau)+\alpha h_{v}(\tau)=-\alpha T$ will do. 

Among these MW fields the critical field can be defined as the one
that minimizes the power injected in the system. The latter is proportional
to the squared intensity of the MW field, i.e. $p_{i}(\tau)=\mathbf{h}^{2}(\tau)=h_{m}^{2}(\tau)+h_{u}^{2}(\tau)+h_{v}^{2}(\tau)$.
Using the method of Lagrange multipliers we define the functional
\begin{equation}
L\left[h_{m}(\tau),\, h_{u}(\tau),\, h_{v}(\tau),\,\lambda(\tau)\right]=p_{i}^{2}(\tau)-\lambda(\tau)\frac{d\mathcal{E}}{d\tau}.\label{eq: calculation bcrit}
\end{equation}

Assuming that $dh_{m}(\tau)/d\tau$ does not depend explicitely on
$h_{m}(\tau),$ the minimization of this functional leads to
\begin{eqnarray}
h_{m}(\tau) & = & 0,\nonumber \\
h_{u}(\tau) & = & -\frac{\alpha}{1+\alpha^{2}}\, T,\label{eq: bcrit components}\\
h_{v}(\tau) & = & \frac{\alpha^{2}}{1+\alpha^{2}}\, T.\nonumber 
\end{eqnarray}

The critical MW field thus reads
\begin{equation}
\mathbf{h}^{\mbox{crit}}(\tau)=\frac{\alpha}{1+\alpha^{2}}\left[-\mathbf{T}+\alpha\,\mathbf{m}\times\mathbf{T}\right]\label{eq: bcrit}
\end{equation}

and the injected power is then
\begin{equation}
p_{i}^{\mbox{crit}}(\tau)=\frac{\alpha^{2}}{1+\alpha^{2}}\, T^{2}.\label{eq: pcrit(t)}
\end{equation}

In the presence of this MW field the magnetization precesses around
the equilibrium position with a constant angle. This critical field
represents a lower limit. Indeed, if the injected power is smaller
than $p_{i}^{\mbox{crit}}(\tau)$ the magnetization goes back to the
initial equilibrium position.

\subsubsection{MW field minimizing the total injected energy: optimal field \label{sub: injected energy}}

In order to minimize the total injected energy we have to make a few
preliminary assumptions concerning the shape of the MW field. Considering
the result of the previous section we limit our search to the family
of MW fields defined by 
\begin{eqnarray*}
h_{m}(\tau) & = & \mbox{\ensuremath{\beta}}_{m}\, T\\
h_{u}(\tau) & = & \mbox{\ensuremath{\beta}}_{u}\, T\\
h_{v}(\tau) & = & \beta_{v}\, T
\end{eqnarray*}
where $\beta_{m}$, $\beta_{u}$ and $\beta_{v}$ are constant parameters.
In the presence of such a MW field the energy variation reads 
\begin{eqnarray}
\frac{d\mathcal{E}}{d\tau} & = & \frac{-\alpha-\beta_{u}+\alpha\beta_{v}}{1+\alpha^{2}}\, T^{2}-\frac{dh_{m}(\tau)}{d\tau}.\label{eq: energy variation with the family of MW fields}
\end{eqnarray}

The total energy injected to the system can be defined as $E=\int_{\tau_{i}}^{\tau_{f}}\mathbf{h}^{2}(\tau)dt$.
Therefore,
\begin{eqnarray}
E & = & \int_{\tau_{i}}^{\tau_{f}}\left(\beta_{m}^{2}+\beta_{u}^{2}+\beta_{v}^{2}\right)\, T^{2}d\tau\label{eq: Emin}\\
 & = & \int_{\tau_{i}}^{\tau f}\left(\beta_{m}^{2}+\beta_{u}^{2}+\beta_{v}^{2}\right)\left[\frac{1+\alpha^{2}}{-\alpha-\beta_{u}+\alpha\beta_{v}}\left(\frac{d\mathcal{E}}{d\tau}+\frac{dh_{m}(\tau)}{d\tau}\right)\right]d\tau\nonumber \\
 & = & \frac{\left(1+\alpha^{2}\right)\left(\beta_{m}^{2}+\beta_{u}^{2}+\beta_{v}^{2}\right)}{-\alpha-\beta_{u}+\alpha\beta_{v}}\left[\mathcal{E}(\tau_{f})-\mathcal{E}(\tau_{i})+h_{m}(\tau_{f})-h_{m}(\tau_{i})\right].\nonumber 
\end{eqnarray}

Hence, the energy is minimal if $\beta_{m}=0$, $\beta_{u}=-2\alpha/\left(1+\alpha^{2}\right)$
and $\beta_{v}=2\alpha^{2}/\left(1+\alpha^{2}\right)$. Consequently,
the optimal MW field is 
\begin{equation}
\mathbf{h}^{\mbox{opt}}(\tau)=\frac{2\alpha}{1+\alpha^{2}}\left[-\mathbf{T}+\alpha\mathbf{m}\times\mathbf{T}\right]=2\,\mathbf{h}^{\mbox{crit}}(\tau).\label{eq: optimal MW field}
\end{equation}

The optimal field is twice the critical field determined previously,
see Eq. (\ref{eq: bcrit}). It is orthogonal to the magnetization
at any time and its magnitude reads
\begin{equation}
\left\Vert \mathbf{h}^{\mbox{opt}}(\tau)\right\Vert =\frac{2\alpha}{\sqrt{1+\alpha^{2}}}\, T.\label{eq: optimal MW field magnitude}
\end{equation}

The total injected energy is then 
\begin{equation}
E=4\alpha\,\left[\mathcal{E}(\tau_{f})-\mathcal{E}(\tau_{i})\right].\label{eq: Emin 2}
\end{equation}

According to these results, both the optimal field magnitude and the
total injected energy increase with damping. This confirms the fact
that the MW field must compensate for the effects of damping so as
to induce switching.

\subsubsection{Trajectory of the magnetization in presence of the optimal MW field\label{sub: trajectory}}

In order to check whether the optimal MW field obtained in the previous
section induces switching of the magnetization as required, we now
investigate the time trajectory of the magnetization. In the presence
of this field, Eqs. (\ref{eq: DLLE in mobile basis}) and (\ref{eq: dwdt in mobile basis})
respectively become 
\begin{eqnarray}
\left(1+\alpha^{2}\right)\frac{d\mathbf{m}}{d\tau} & = & -\mathbf{T}+\alpha\mathbf{m}\times\mathbf{T},\label{eq: DLLE with bopt}
\end{eqnarray}
\begin{equation}
\frac{d\mathcal{E}}{d\tau}=\frac{\alpha}{1+\alpha^{2}}\, T^{2}.\label{eq: dwdt with bopt}
\end{equation}

The first equation is similar to the Landau-Lifshitz-Gilbert equation
but with a negative damping parameter: it describes an amplified precession.
The precession frequency is equal to the proper frequency of the magnetization.
At any time the MW field is proportional to the derivative of the
magnetization: $\mathbf{h}^{\mbox{opt}}(\tau)=2\alpha\, d\mathbf{m}/d\tau$.
This is in agreement with the results of Sun et al. \cite{sun06bis}. 

At the minima and saddle points the effective field $\mathbf{h}_{\mbox{eff}}$
is parallel to the magnetization so that $\mathbf{T}=\mathbf{0}$.
Therefore, both the derivative of the magnetization and the MW field
vanish. Consequently, the optimal MW field can only induce the motion
of the magnetization from an initial state $\mathbf{m}_{i}$ close
to an energy minimum, to a final state $\mathbf{m}_{f}$ close to
a saddle point. A small amount of energy must thus be added (i) before
the MW field pulse, to drag the magnetization away from the minimum
and (ii) after the pulse, to cross the saddle point. The nature of
this additionnal energy will be further discussed later on. Beyond
the saddle point, the damping takes up to lead the magnetization down
to the second energy minimum. If the energy landscape is complex with
several barriers successive pulses might then be necessary to induce
switching. 

At both the initial and the final states the MW field is close to
zero. The difference $\mathcal{E}(\tau_{f})-\mathcal{E}(\tau_{i})$
is thus close to the static energy barrier between the saddle point
and the initial state $\triangle\mathcal{E}_{0}\equiv\mathcal{E}_{0}(\tau_{f})-\mathcal{E}_{0}(\tau_{i})$
and Eq. (\ref{eq: Emin 2}) becomes 
\begin{equation}
E=4\alpha\,\triangle\mathcal{E}_{0}.\label{eq: Emin3}
\end{equation}

The total injected energy is therefore proportional to the energy
barrier to be overcome. Hence, if the static field is close to the
switching field, a very weak MW field can induce switching.

\subsection{Uniaxial anisotropy and longitudinal static field\label{sub: uniaxial anisotropy and longitudinal field}}

In this section we study the trajectory of the magnetization in the
presence of the optimal MW field for a nanoparticle with uniaxial
anisotropy and a longitudinal static field. 

We consider a nanomagnet with uniaxial anisotropy with easy axis in
the $z$ direction. The anisotropy energy density is then $\mathcal{E}_{\mbox{an}}(m_{z})=-m_{z}^{2}/2$.
The static field is applied in the $(-z)$ direction with a magnitude
$0\leq h_{0}<1$. The magnetization is initially close to the metastable
minimum and its $z$ component is thus $m_{0}\equiv m_{z}(\tau_{i}=0)\approx1$.
The static energy of the system is 
\begin{equation}
\mathcal{E}_{0}(m_{z})=-\frac{m_{z}^{2}}{2}+h_{0}\, m_{z}.\label{eq: energy barrier long field}
\end{equation}

The saddle point then corresponds to $m_{z}=h_{0}$, so the static
energy barrier between the latter and the initial metastable state
is $\triangle\mathcal{E}_{0}=\left(1-h_{0}\right)^{2}/2$. For this
system, the effective field reads $\mathbf{h}_{\mbox{eff}}=-h_{0}+m_{z}\mathbf{e}_{z}.$
Projecting Eq. (\ref{eq: DLLE with bopt}) onto the $z$ axis then
yields 
\begin{eqnarray}
\frac{dm_{z}}{d\tau} & = & \frac{\alpha}{1+\alpha^{2}}\left(\mathbf{m}\times\mathbf{T}\right)\cdot\mathbf{e}_{z}=-\frac{\alpha}{1+\alpha^{2}}\,\left(-h_{0}+m_{z}\right)\left(1-m_{z}^{2}\right).\label{eq: DLLE along z}
\end{eqnarray}

In order to simplify the expressions we introduce the integral
\begin{eqnarray}
I(m_{z}) & = & \int_{0}^{m_{z}}\frac{-du}{\left(-h_{0}+u\right)\left(1-u^{2}\right)}=\frac{1}{2\left(1-x^{2}\right)}\ln\left[\frac{\left(1+m_{z}\right)^{1-h_{0}}\left(1-m_{z}\right)^{1+h_{0}}}{\left(-h_{0}+m_{z}\right)^{2}}\right]+C^{te}.\label{eq: I(mz)}
\end{eqnarray}

Solving Eq. (\ref{eq: DLLE along z}) with the initial condition $m_{z}(t=0)=m_{0}$
leads to the equation
\begin{equation}
I(m_{z})-I(m_{0})=\frac{\alpha}{1+\alpha^{2}}\,\tau.\label{eq: z traj}
\end{equation}

\begin{figure}[h]
\begin{centering}
\includegraphics[width=0.8\columnwidth]{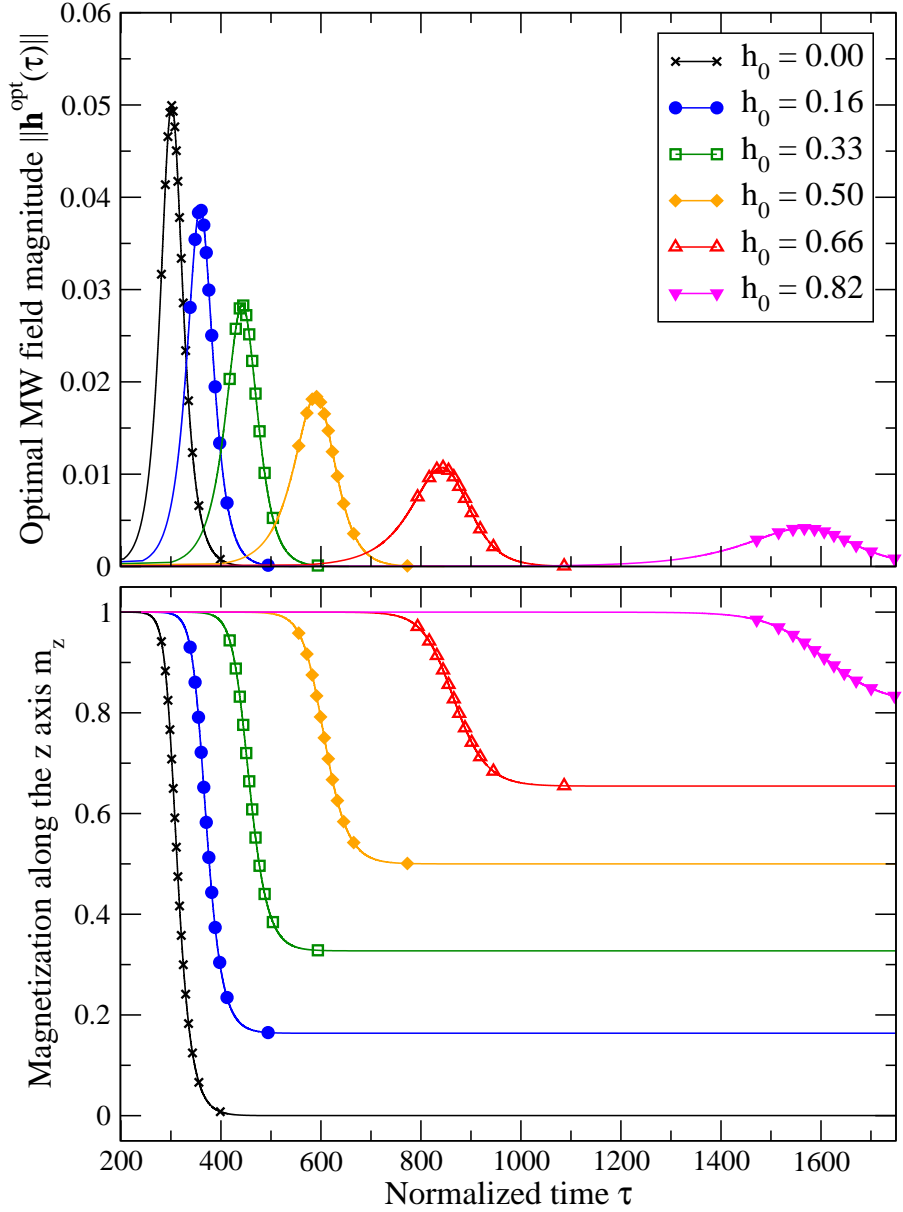}
\par\end{centering}

\caption{(Color online) Optimal MW field intensity $\left\Vert \mathbf{h}^{\mbox{opt}}(\tau)\right\Vert $
(upper panel) and trajectory of the magnetization $m_{z}(\tau)$ (lower
panel) for a longitudinal static field with magnitude $h_{0}$. Parameters:
$\alpha=0.05$, $m_{0}=0.99998$. }

\label{fig: ANA - MW field and spin traj vs reduced time} 
\end{figure}

This equation can be analytically solved for $m_{z}$ only if $h_{0}=0$
(no static field). Otherwise, the evolution of $m_{z}$ with time
can be obtained numerically, see Fig. \ref{fig: ANA - MW field and spin traj vs reduced time}.
As predicted previously, for long times the magnetization goes towards
to the saddle point but never reaches it since $I(m_{z})$ diverges
for $m_{z}=h_{0}$. 

From Eq. (\ref{eq: optimal MW field magnitude}) we can express the
optimal MW field intensity in terms of $m_{z}$ as follows 
\begin{eqnarray}
\left\Vert \mathbf{h}^{\mbox{opt}}(m_{z})\right\Vert  & = & \frac{2\alpha}{\sqrt{1+\alpha^{2}}}(-h_{0}+m_{z})\sqrt{1-m_{z}^{2}}.\label{eq: bopt intensity}
\end{eqnarray}

The time evolution of the MW field intensity is plotted in Fig. \ref{fig: ANA - MW field and spin traj vs reduced time}.
We note that the pulses follow neither a Gaussian nor a Lorentzian
function. The peak intensity is reached for $m_{z}=\frac{1}{4}\left(h_{0}+\sqrt{8+h_{0}^{2}}\right)$
and is given by
\begin{equation}
h_{\mbox{max}}^{\mbox{opt}}=\frac{\alpha}{2\sqrt{1+\alpha^{2}}}\left(-3h_{0}+\sqrt{8+h_{0}^{2}}\right)\sqrt{1-\frac{\left(h_{0}+\sqrt{8+h_{0}^{2}}\right)^{2}}{16}}.\label{eq: bmax}
\end{equation}

\begin{figure}[h]
\begin{centering}
\includegraphics[angle=270,width=0.8\columnwidth]{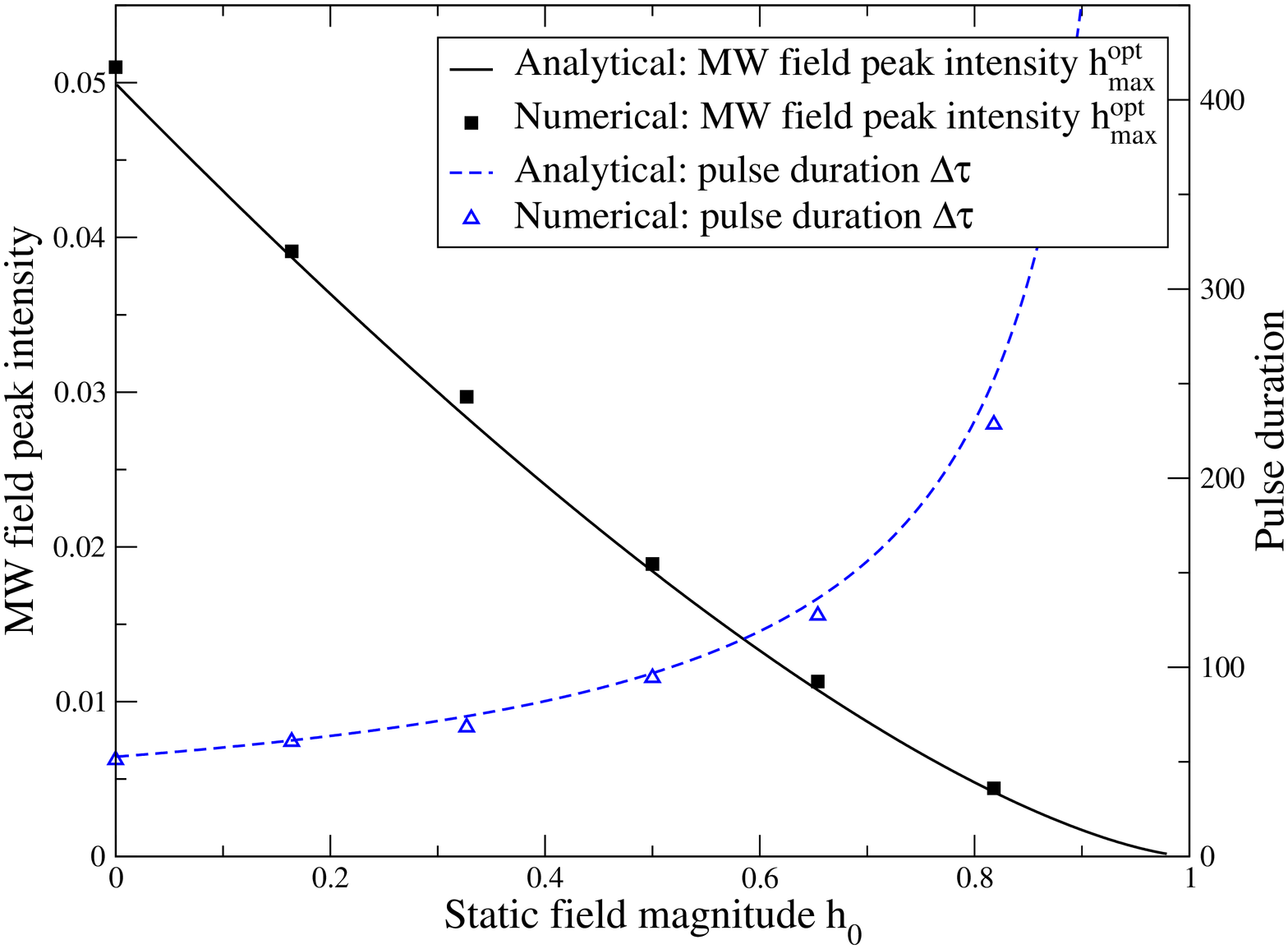}
\par\end{centering}

\caption{(Color online) Maximal peak intensity and peak duration of the optimal
MW field for varying magnitude of the static field $h_{0}$.}
\label{fig: magnitude}
\end{figure}

From this, we can see that the peak intensity $h_{\mbox{max}}^{\mbox{opt}}$
decreases with $h_{0}$ (see Fig. \ref{fig: magnitude}). Indeed,
for higher magnitudes of the static field, the energy barrier $\triangle\mathcal{E}_{0}$
between the metastable state and the saddle point is lower, so that
a lower energy is needed to reach the saddle point. Since $0\leq h_{0}<1$
the peak intensity is limited as follows 
\begin{equation}
h_{\mbox{max}}^{\mbox{opt}}<\frac{\sqrt{2}\alpha}{\sqrt{1+\alpha^{2}}}.\label{eq: bmax limited}
\end{equation}

Hence, for low values of the damping parameter $\alpha$, the intensity
of the optimal MW field is small. This fully confirms the results
of our numerical study \cite{barros11}. Using Eqs. (\ref{eq: I(mz)})
and (\ref{eq: bopt intensity}), we can also obtain analytically the
pulse duration $\triangle\tau$, defined as the full width at half
maximum 
\begin{equation}
\triangle\tau=\frac{1+\alpha^{2}}{\alpha}g\left(h_{0}\right),\label{eq: deltat}
\end{equation}
where $g\left(h_{0}\right)$ is a cumbersome function of $h_{0}$.
This characteristic time increases with $h_{0}$ (see Fig. \ref{fig: magnitude}).
For high values of $h_{0}$, the switching will thus require a very
low field but a very long time, as can be seen in Fig. \ref{fig: ANA - MW field and spin traj vs reduced time}.
Moreover, the characteristic time decreases with damping. 

The area below the curves $\left\Vert \mathbf{h}^{\mbox{opt}}(\tau)\right\Vert $
is
\begin{eqnarray}
A & = & \int_{t=0}^{t=\infty}\left\Vert \mathbf{h}^{\mbox{opt}}(\tau)\right\Vert d\tau=\int_{m_{z}=m_{0}}^{m_{z}=h_{0}}\left\Vert \mathbf{h}^{\mbox{opt}}(m_{z})\right\Vert \frac{d\tau}{dm_{z}}dm_{z}\label{eq: area below b(t)}\\
 & = & 2\sqrt{1+\alpha^{2}}\left[\arccos\left(h_{0}\right)-\arccos\left(m_{0}\right)\right]=2\sqrt{1+\alpha^{2}}\left(\theta_{f}-\theta_{i}\right).\nonumber 
\end{eqnarray}
 where $\theta_{i}$ and $\theta_{f}$ are the polar angles of the
magnetization respectively at the initial state and saddle point.
This area is thus proportional to the ``angular distance'' that
the magnetization must cross to reach the saddle point. For increasing
values of the static field magnitude $h_{0}$, this area decreases
since the saddle point comes closer to the initial state. 

The $z$ component of $\mathbf{h}^{\mbox{opt}}$, given by Eq. (\ref{eq: optimal MW field}),
is
\begin{eqnarray}
h_{z}^{\mbox{opt}}(m_{z}) & = & -\frac{2\alpha^{2}}{1+\alpha^{2}}(-h_{0}+m_{z})\left(1-m_{z}^{2}\right).\label{eq: bopt z component}
\end{eqnarray}

From Eq. (\ref{eq: bopt intensity}) we can see that the ratio $\left|h_{z}^{\mbox{opt}}(m_{z})\right|/\left\Vert \mathbf{h}^{\mbox{opt}}(m_{z})\right\Vert $
is $\frac{\alpha}{\sqrt{1+\alpha^{2}}}\sqrt{\left(1-m_{z}^{2}\right)}$,
whose upper limit is $\frac{\alpha}{\sqrt{1+\alpha^{2}}}$. Consequently,
for small values of the damping parameter $\alpha$, the component
of the time-dependent field along the anisotropy easy axis can be
neglected and the optimal field lies in the $xy$ plane.

We now define the precession phase of the magnetization as $\varphi(\tau)=\arctan\left(m_{y}(\tau)/m_{x}(\tau)\right)$.
Projecting Eq. (\ref{eq: DLLE with bopt}) on the $x$ and $y$ axes
leads to the relation
\begin{equation}
\omega(\tau)=\frac{d\varphi}{d\tau}=\frac{\left\Vert \mathbf{h}_{\mbox{eff}}\right\Vert }{1+\alpha^{2}}=\frac{-h_{0}+m_{z}(\tau)}{1+\alpha^{2}}.\label{eq: omega(t)}
\end{equation}

This precession frequency is equal to the proper frequency of the
magnetization, obtained by solving Eq. \ref{eq: DLLE} in the absence
of a MW field. At the initial state, the precession frequency is close
to the FMR frequency $\omega_{FMR}=\left(1-h_{0}\right)/\left(1+\alpha^{2}\right)$.
It then decreases towards zero, following the curvature of the energy
well. 

For small values of $\alpha$, since the optimal field lies in the
$xy$ plane as shown previously, its phase can be defined by $\tilde{\varphi}(t)=\arctan\left(h_{y}^{\mbox{opt}}(t)/h_{x}^{\mbox{opt}}(t)\right)$.
It can be shown that
\begin{equation}
\tan\tilde{\varphi}(\tau)=\frac{m_{x}(\tau)+\alpha m_{y}(\tau)\, m_{z}(\tau)}{-m_{y}(\tau)+\alpha m_{x}(\tau)\, m_{z}(\tau)}\approx-\frac{m_{x}(\tau)}{m_{y}(\tau)}=\cot\varphi(\tau).\label{eq: bopt phase}
\end{equation}

This implies that the time-dependent field and the magnetization are
synchronized with $\tilde{\varphi}(\tau)\approx\varphi(\tau)+\pi/2$.
Hence, the frequency of the time-dependent field is equal to the proper
precession frequency of the magnetization.

Finally, using Eqs. (\ref{eq: DLLE along z}) and (\ref{eq: bopt intensity}),
the total injected energy can be computed. As shown previously in
Eq. (\ref{eq: Emin3}), it is proportional to the damping parameter
and to the energy barrier $\triangle\mathcal{E}_{0}$. 
\begin{equation}
E=\int_{0}^{+\infty}\left\Vert \mathbf{h}^{\mbox{opt}}(\tau)\right\Vert ^{2}d\tau=2\alpha\left(h_{0}-1\right)^{2}=4\alpha\triangle\mathcal{E}_{0}.\label{eq: total energy with uniax anis}
\end{equation}

\section{Comparison with the numerical results}

As mentioned earlier, in Ref. \onlinecite{barros11} we developed
a numerical method based on the theory of optimal control to determine
the shape of the optimal MW field. It renders an exact solution for
the MW field that triggers the switching of a nanomagnet with a given
anisotropy and applied field. The method consists in minimizing the
cost functional 
\[
\mathcal{F}\left[\mathbf{m}(\tau),\mathbf{h}(\tau)\right]=\frac{1}{2}\left\Vert \mathbf{m}(\tau_{f})\mathbf{-m}_{f}\right\Vert ^{2}+\frac{\eta}{2}\int\limits _{0}^{\tau_{f}}d\tau\,\mathbf{h}^{2}(\tau)
\]
along the trajectory given by the Landau-Lifshitz-Gilbert equation
(\ref{eq: DLLE}), where $\mathbf{m}_{f}$ is the target magnetization
(stable minimum), $\mathbf{m}(\tau_{f})$ is the magnetization reached
at time $\tau_{f}$ and $\eta$ is a numerical control parameter.
The numerical problem is then solved using the modified conjugate
gradient technique supplemented by a Metropolis algorithm. 

In Ref. \onlinecite{barros11} we restricted the MW field along a
polarization axis to comply with the experimental setup. In the present
study, for a better comparison with the analytical results, the MW
field is allowed to move in three dimensions during the optimization.

Our model system is a particle with uniaxial anisotropy along the
$z$ axis. Unless otherwise specified, the numerical parameters used
in the current study are: initial normalized time $\tau_{i}=0$, final
normalized time $\tau_{f}=800$ (corresponding to a few ns in real
time), number of points $N=15000$, so the sampling time $(\tau_{f}-\tau_{i})/(N-1)$
is about 0.05; damping parameter $\alpha=0.05$; control parameter
$\eta=0.01$.

\subsection{Reference calculation}

\begin{figure}[h]
\begin{centering}
\includegraphics[width=0.8\columnwidth]{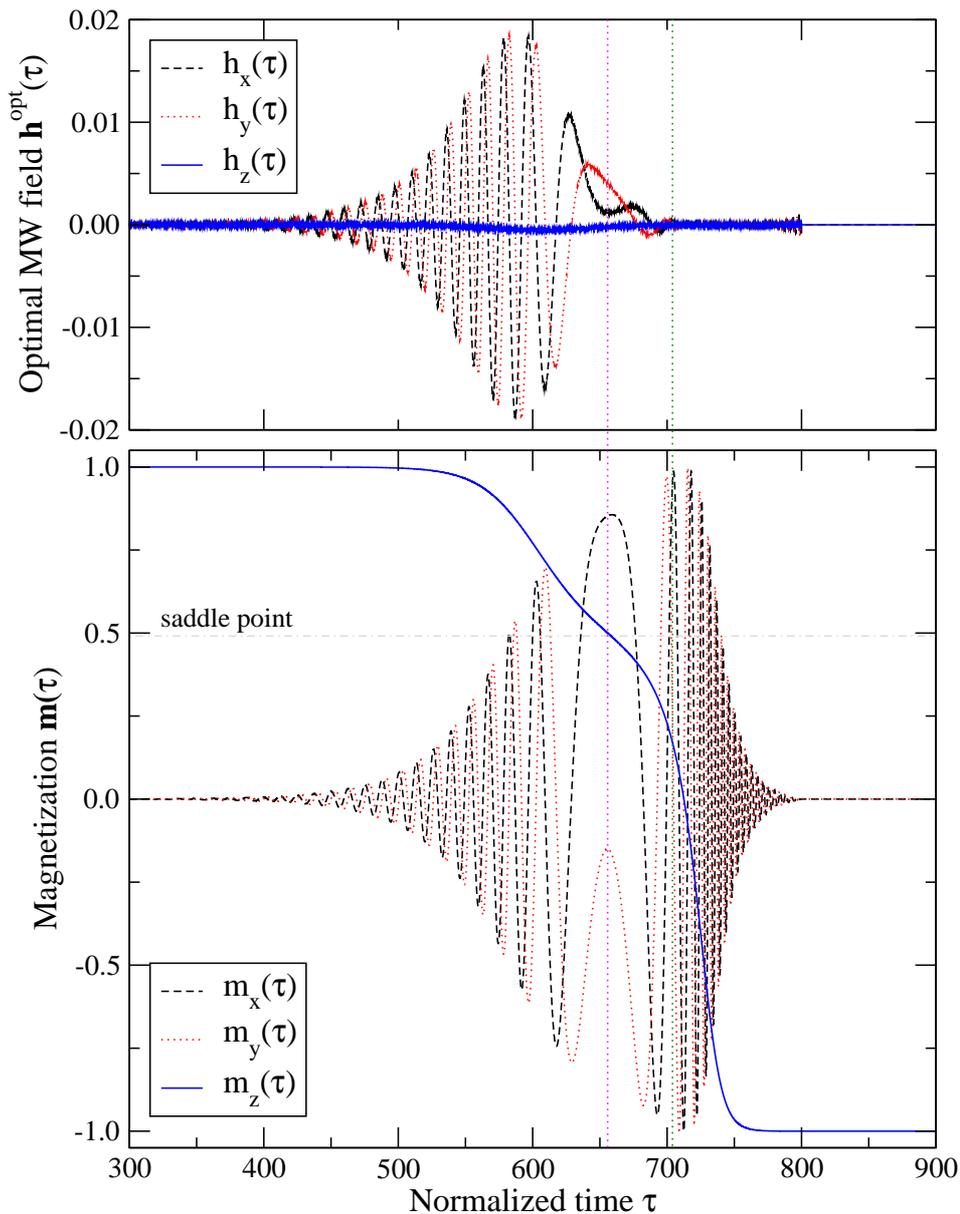}
\par\end{centering}

\caption{(Color online) Numerical results for the reference calculation: optimal
MW field (upper panel) and magnetization time trajectory (lower panel).
The purple and green dotted vertical lines respectively indicate the
crossing of the saddle point and the end of the microwave pulse. }

\label{fig: reference calculation}
\end{figure}

A first numerical optimization was carried out for a static field
with magnitude $h_{0}=0.5$ applied in the ($-z$) direction. The
results are in good agreement with the analytical calculations of
part \ref{sec: Analytic-calculations}. As can be seen in Fig. \ref{fig: reference calculation},
the optimal MW field is modulated both in amplitude and frequency.
It is mainly in the $xy$ plane and its magnitude is small (less than
$0.03$, corresponding to a few mT in real field) as expected since
the damping parameter is small. The pulse starts at about $\tau=350$
and progressively drives the magnetization away from the initial equilibrium
position. The saddle point is reached at about $\tau=650$ (purple
dotted line) but the MW field pulse continues until $\tau=700$ (green
dotted line), which allows the magnetization to cross the saddle point.
Next, the damping takes up to lead the magnetization to the more stable
energy minimum, which is reached at about $\tau=800$. 

\begin{figure}[h]
\begin{centering}
\includegraphics[angle=270,width=0.8\columnwidth]{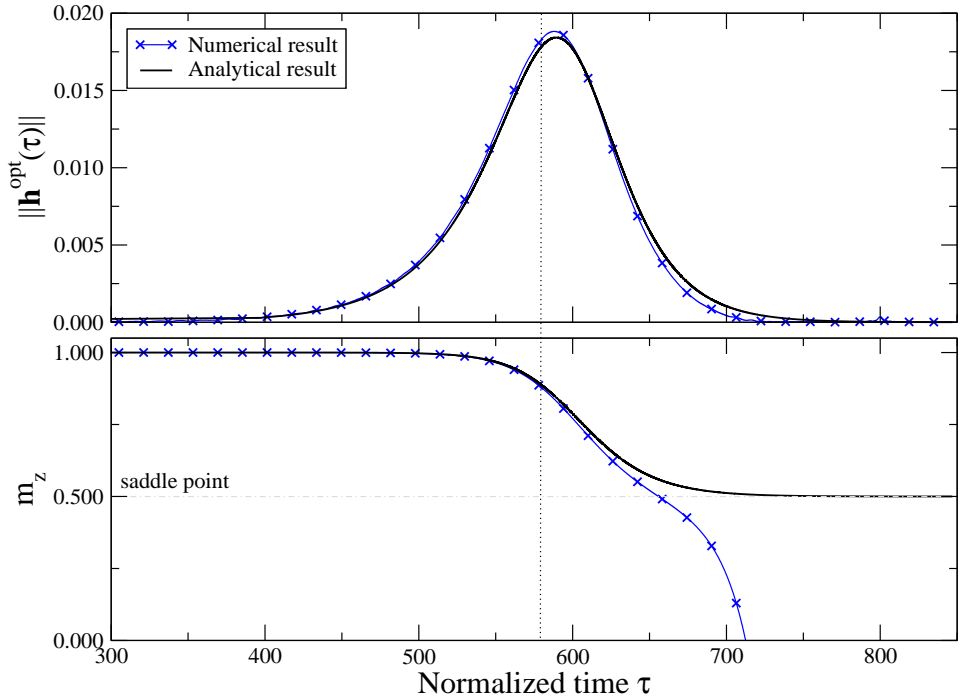}
\par\end{centering}

\caption{(Color online) Comparison between the analytical and numerical results
for the reference calculation: optimal MW field magnitude $\left\Vert \mathbf{h}^{\mbox{opt}}\right\Vert $(higher
panel) and $z$ component of the magnetization (lower panel). }
\label{fig: reference - magnitude and mz}
\end{figure}

Fig. \ref{fig: reference - magnitude and mz} shows that the MW field
intensity obtained numerically is in good agreement with the analytical
result in Eq. (\ref{eq: bopt intensity}). From $\tau\approx570$,
the MW field intensity is slightly higher numerically, which induces
$m_{z}$ to decrease faster and the magnetization to finally cross
the saddle point. The total injected energy obtained numerically $E^{\mbox{num}}=\int_{\tau_{i}}^{\tau_{f}}\mathbf{h}^{2}(\tau)dt\approx0.02548$
is slightly higher than the value predicted analytically $E^{\mbox{an}}=4\alpha\,\triangle\mathcal{E}_{0}=0.02500$.
This confirms that the optimal MW field determined analytically represents
a lower boundary and that a small additionnal energy must be injected
to achieve switching, as noticed previously. Nevertheless, the discrepancy
between the numerical and analytical MW fields is very small, which
corroborates the relevance of the analytical model. 

\begin{figure}[h]
\begin{centering}
\includegraphics[angle=270,width=0.8\columnwidth]{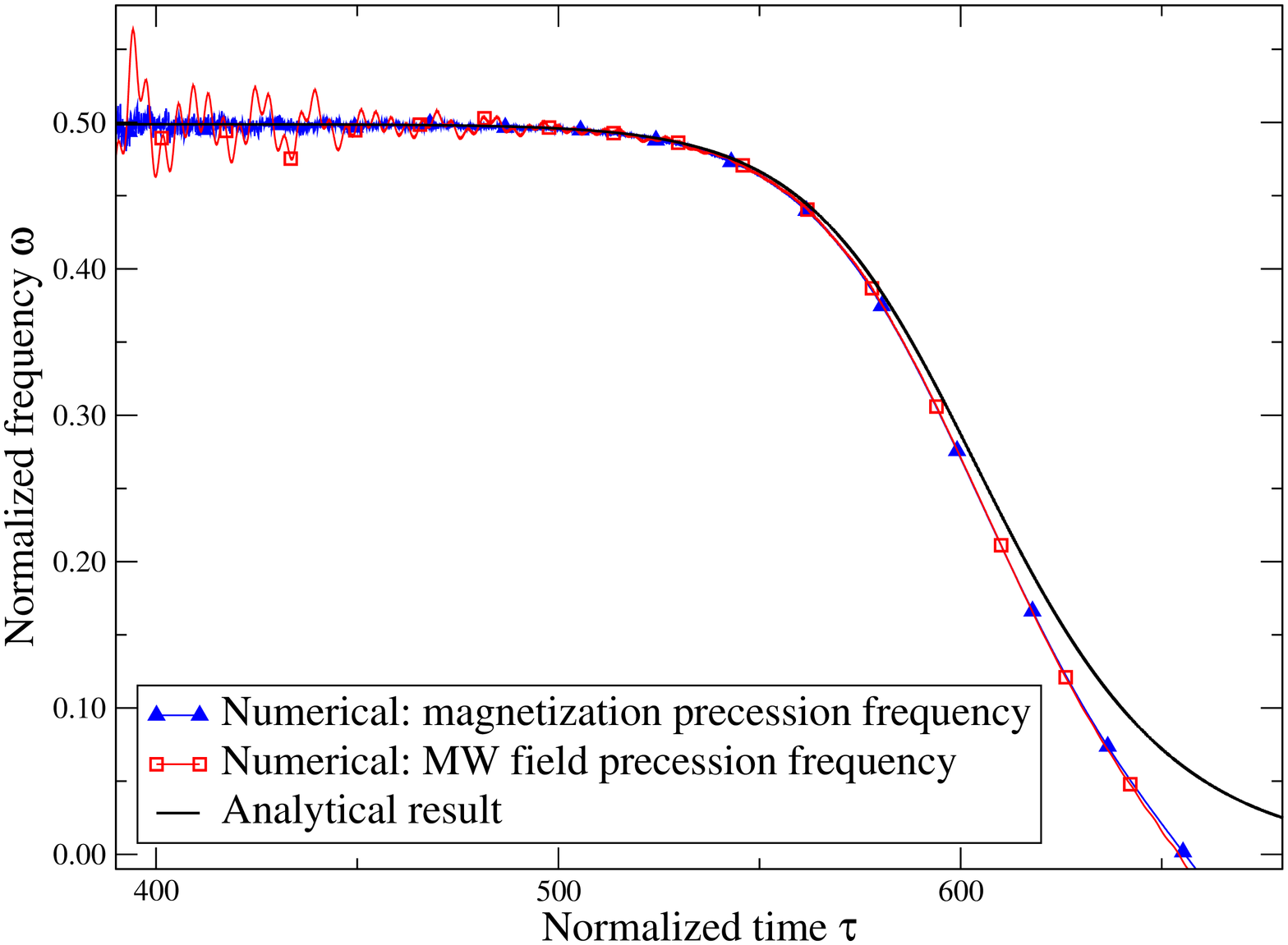}
\par\end{centering}

\caption{(Color online) Comparison between the analytical and numerical results
for the reference calculation: precession frequencies of the magnetization
and MW field. }
\label{fig: reference - frequency}
\end{figure}

Fig. \ref{fig: reference - frequency} confirms that the magnetization
precession and the MW field are synchronized, the initial frequency
being close to the FMR frequency. The time evolution of the frequency
is similar to the evolution of $m_{z}$ (Fig.\ref{fig: reference - magnitude and mz}),
since both values are related by Eq. (\ref{eq: omega(t)}). Consequently,
after $\tau\approx570$ the numerical frequency is lower that the
analytical frequency. 

\subsection{Effect of the static field magnitude and direction}

\begin{figure}[h]
\begin{centering}
\includegraphics[angle=270,width=0.8\columnwidth]{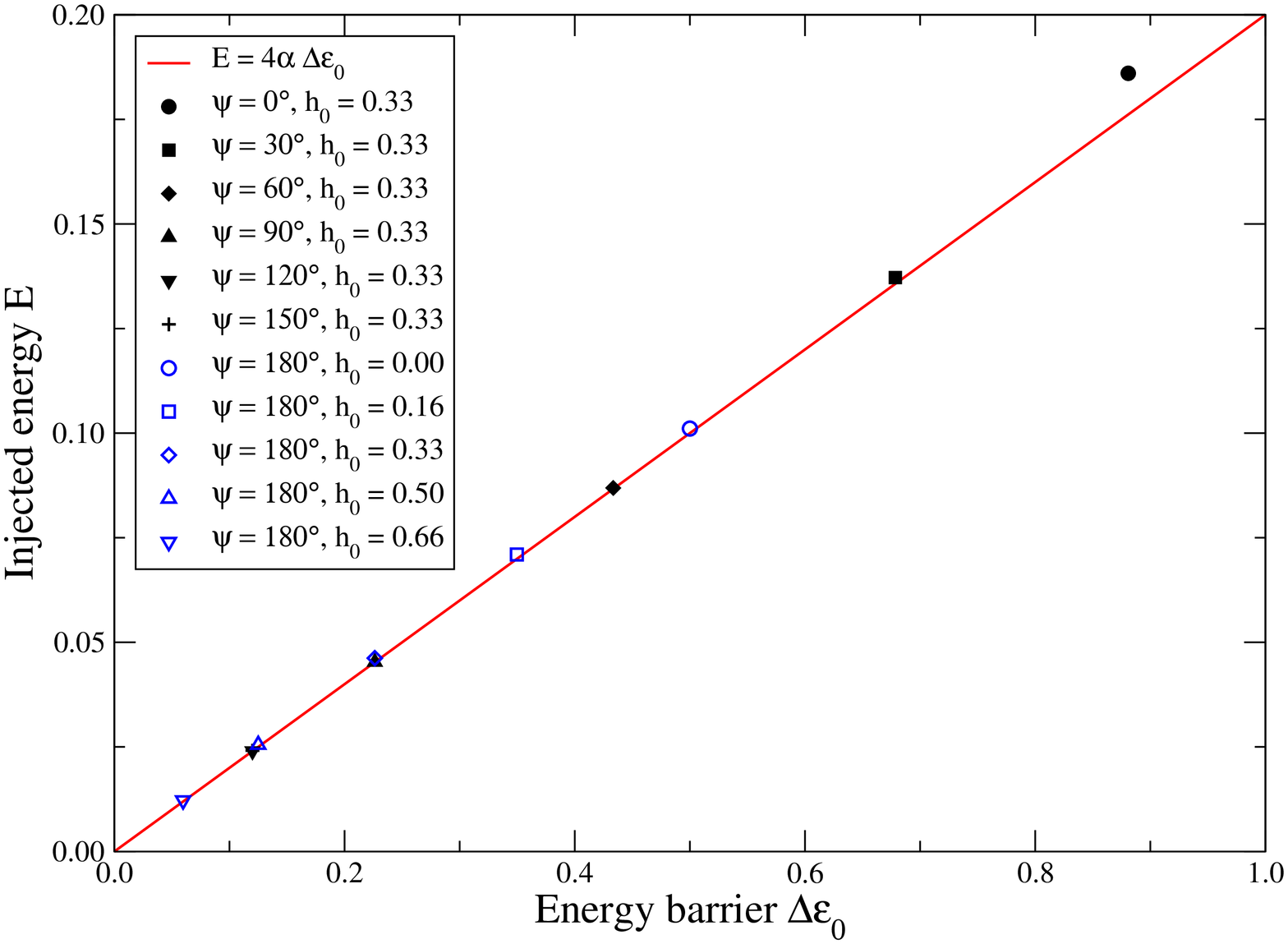}
\par\end{centering}

\caption{(Color online) Total injected energy $E$ with respect to the static
energy barrier $\triangle\mathcal{E}_{0}$ for varying magnitude and
orientation of the static field $\mathbf{h}_{0}$. $\psi$ is the
angle between the $z$ axis (anisotropy axis) and the static field.
\label{fig: E vs DW0}}
\end{figure}

The MW field $\mathbf{h}(\tau)$ has been optimized numerically for
several magnitudes and orientations of the static field $\mathbf{h}_{0}$.
For each configuration, the energy barrier $\triangle\mathcal{E}_{0}$
and the total injected energy $E=\int_{\tau_{i}}^{\tau_{f}}\left\Vert \mathbf{h}^{\mbox{opt}}(\tau)\right\Vert ^{2}d\tau$
have been computed numerically, and are reported in Fig. \ref{fig: E vs DW0}.
As shown in Eq. (\ref{eq: Emin3}), the injected energy is found to
be proportional to the energy barrier and to $4\alpha$. 

In the case of a static field applied along ($-z$), the shape of
the pulse can be directly compared with the analytical results of
sec. \ref{sub: uniaxial anisotropy and longitudinal field}, see Fig.
\ref{fig: magnitude}. The numerical and analytical results are in
good agreement. As predicted analytically, the maximal peak intensity
$h_{\mbox{max}}^{\mbox{opt}}$ decreases and the pulse duration $\triangle\tau$
increases rapidly when the magnitude of the static field $h_{0}$
increases.

\subsection{Effect of damping}

\begin{figure}[h]
\begin{centering}
\includegraphics[angle=270,width=0.8\columnwidth]{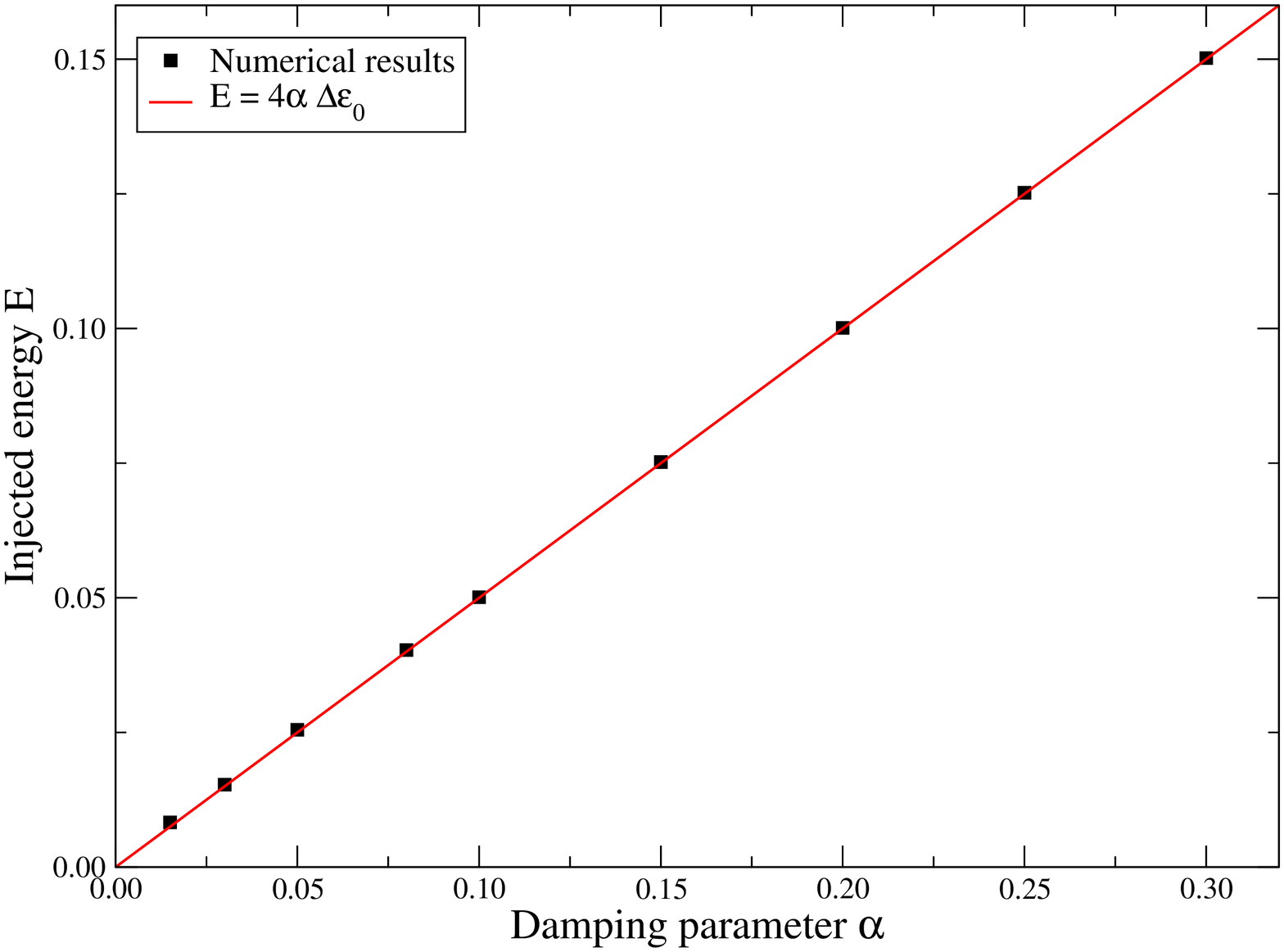}
\par\end{centering}

\caption{(Color online) Total injected energy $E$ with respect to the damping
parameter $\alpha$. \label{fig: E vs alpha}}
\end{figure}

As predicted by Eqs. (\ref{eq: Emin3}) and (\ref{eq: total energy with uniax anis}),
for a given static energy barrier $\triangle\mathcal{E}_{0}$, the
injected energy is proportional to the damping parameter $\alpha$.
This has been checked numerically by varying the damping parameter
from 0.015 to 0.30 (Fig. \ref{fig: E vs alpha}). In these calculations,
the static field $\mathbf{h}_{0}$ is applied in the ($-z$) direction
with the magnitude $h_{0}=0.5$. 

\begin{figure}[h]
\begin{centering}
\includegraphics[angle=270,width=0.8\columnwidth]{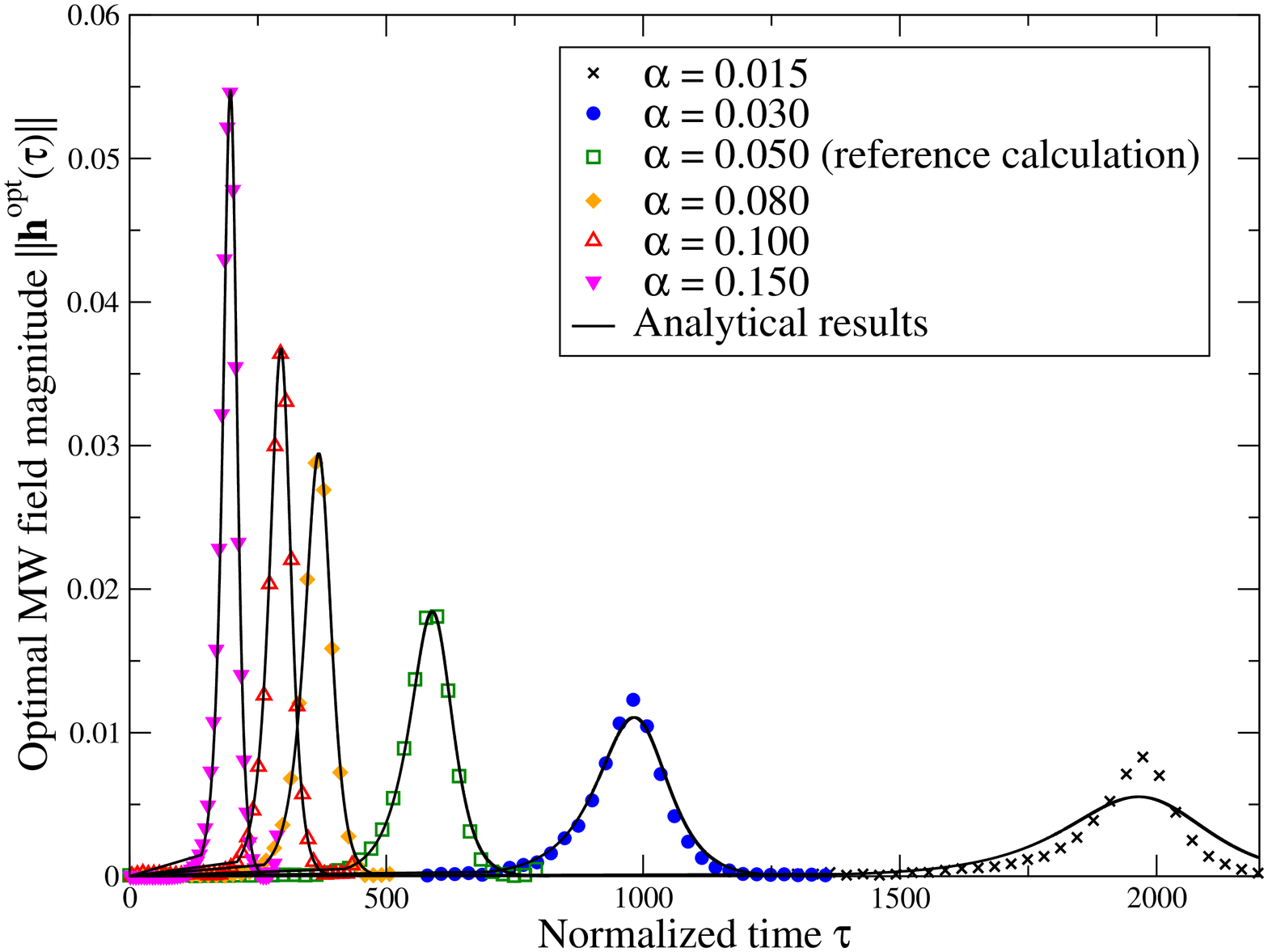}
\par\end{centering}

\caption{(Color online) Optimal MW field pulse for several values of the damping
parameter $\alpha$. For $\alpha=0.015$, the final time has been
increased ($\tau_{f}=1600$) without changing the sampling time. }
\label{fig: damping}
\end{figure}

Fig. \ref{fig: damping} shows that for low values of $\alpha$, the
pulses height decreases but their duration increases, in agreement
with Eqs. (\ref{eq: bmax}) and (\ref{eq: deltat}). For an undamped
system, the switching should thus be infinitely long, so our analytical
and numerical methods are not adapted to describe such a system. 

As can be observed, for very low values of the damping parameter $\alpha$,
a discrepancy between the analytical calculations and the numerical
optimization is observed. Indeed, since the optimal peak duration
becomes very long, the number of numerical points $N$ must be increased,
so the conjugate gradient algorithm becomes less efficient. However,
as can be seen in Fig. \ref{fig: E vs alpha}, this discrepancy has
a negligeable effect on the injected energy. 

\section{Conclusion}

We analytically determined the optimal microwave field that allows
for the switching of the magnetization of a monodomain nanoparticle
with uniaxial anisotropy while minimizing the injected energy. This
study provides a clear interpretation of the results obtained numerically
using the optimal control theory \cite{barros11}, especially the
simple dependence of the pulse on the damping parameter. 

Our results confirm that the optimal MW field is modulated both in
amplitude and frequency, since it is directly proportional to the
derivative of the magnetization. It drives the magnetization from
an initial state close to the initial minimum to a final state close
to a saddle point. The time trajectory can then be described as an
amplified precession. 

In order to cross the saddle point, a small additionnal energy must
be injected into the system. Our numerical results show that this
energy can be added by slightly increasing the MW field intensity.
In reality any source of noise, such as thermal fluctuations, may
suffice to induce the saddle point crossing. Subsequently, the damping
induces the relaxation to the final state. We find that the injected
energy is proportional to the damping parameter and to the energy
barrier between the initial state and the saddle point. For typical
values of the damping parameter ($\alpha<1$), a weak MW field of
a few mT is thus sufficient to induce switching.

For a nanomagnet with uniaxial anisotropy placed in a longitudinal
static field, the shape of the MW field pulse has been obtained analytically.
We have shown that the optimal MW field pulse becomes lower but more
spread when the damping decreases. 

In the case of more complex energy landscapes (with biaxial or cubic
anisotropies) the switching is likely to be triggered by a succession
of MW field pulses. This hypothesis will be later tested numerically.
This study could then be extended to small nanomagnets where surface
effects can not be neglected using the effective one-spin model (EOSP)
\cite{garanin03,kachkachi06,yanes07}. Moreover, the influence of
temperature on the optimization could be investigated numerically
using the Langevin approach which introduces the temperature dependence
through an additionnal stochastic field \cite{brown63,ragusa08}.
In particular, we intend to investigate the conditions under which
the thermal fluctuations can favour the switching by assisting the
magnetization in crossing the saddle point. 

The optimal MW fields that we have found have an amplitude and a frequency
which vary slowly and can be reproduced experimentally using a function
generator. Consequently, our theoretical results could be used to
probe the damping parameter and assess the role of surface effects
in real nanoparticles. The dependence of the MW field on the energy
landscape might be used to address directly a given nanoparticle in
a polydisperse assembly.

\section*{ACKNOWLEDGMENTS}

We are grateful to our collaborators E. Bonet, C. Thirion (Institut
Néel, Grenoble, France) and V. Dupuis (LPMCN, Lyon, France) for instructive
discussions on the microwave-assisted switching of isolated nanoclusters.
This work has been partly funded by the collaborative program PNANO
ANR-08-P147-36 of the French Ministry.\bibliographystyle{apsrev}
\bibliography{mas}

\end{document}